\documentstyle[twoside,fleqn]{article}
\makeatletter                                            %KT
\@ifundefined{epsfbox}{\@input{epsf.sty}}{\relax}        %KT
\def\plotone#1{\centering \leavevmode                    %KT
\epsfxsize=\columnwidth \epsfbox{#1}}                    %KT
\def\plotone_reduction#1#2{\centering \leavevmode        %KT
\epsfxsize=#2\columnwidth \epsfbox{#1}}                  %KT
\makeatother

\begin{document}
\input epsf.sty
\pagestyle{myheadings}
\markboth{\hfill{\protect\footnotesize\rm Ledenev V.G., Karlicky M., Yan Y., Fu Q.}\hfill}
{\hfill{\protect\footnotesize\rm On an estimation of the coronal magnetic field strength}\hfill}

\noindent {\bf ASTROPHYSICS REPORTS} \hfill No.~4, 
{\it Page 00--00}

\noindent {\small Publ. Beijing Astronomical Observatory} \hfill December
2001

\noindent $\hrulefill$

\medskip
\bigskip
\begin{center}
\begin{LARGE}
On an estimation of the coronal magnetic field strength from spectrographic observations in the microwave range
\end{LARGE}

\bigskip
{\large LEDENEV$^{1,3,4}$ Vladimir, KARLICKY$^{2}$ Marian, YAN$^{3,4}$ Yihua, FU$^{3,4}$ Qijun}

\bigskip
\medskip
{\footnotesize $^{1}$Institute of Solar-Terrestrial Physics, Irkutsk 664033, P.O.Box 4026, Russia}

{\footnotesize$ ^{2}$Astronomical Institute, CZ-25165, Ondrejov, Czech Republic}

{\footnotesize$ ^{3}$Beijing Astronomical Observatory, Chinese Academy of 
Sciences, Beijing 100080, China}

{\footnotesize$ ^{4}$National Astronomical Observatories, Chinese Academy of 
Sciences, Beijing 100012, China}

\bigskip
{\small{\bf ABSTRACT}}
\end{center}

{\small
\begin{quotation}
Observations of the solar radio emission in the microwave frequency range show fine structures consisting of a number of the almost parallel narrow frequency bands. We interpret these bands as the cyclotron harmonics plasma emission. This emission is generated by the anisotropic electron beam, which excites longitudinal waves on the normal Doppler effect resonance. Then the longitudinal waves convert into the radio emission on the second harmonic of the longitudinal waves frequency and sometimes into the fundamental one. Estimations of the magnetic field strength made on the basis of such a model give the values of the magnetic field in the microwave burst sources as $\sim$ 100-200 G. The estimations of density variations are also given. 

\end{quotation}


\section[]{Introduction}

There are many works where the estimations of the coronal magnetic field strength are made using solar radio emission observations (see, for example, Dulk and McLean, 1977; Mann et al., 1987; Gelfreikh, 1998; Ledenev and Messerotti, 1999). Observations of the fine structures of the microwave bursts, which were made recently (Huang and Fu, 1999; Izliker and Benz, 1994; Karlicky and Jiricka, 1995, Karlicky et al., 1996) open new possibilities for a determination of the coronal magnetic field strength. Bursts of the solar emission with the superimposed fine structures are observed in the microwave range frequently. It is natural to suppose that properties of these narrowband structures with the bandwidth about few percents are determined by the coronal magnetic field, because the dispersion characteristics of the plasma become complicated significantly in the presence of the magnetic field. Sometimes these fine structures consist of a number of the narrow bands, which are almost parallel each other. It is natural to suppose that these bands are the cyclotron harmonics. But if we take the distance between adjacent frequency bands as the cyclotron harmonic frequency we get the numbers of harmonics of the order of some tens. It is unlikely to generate an intensive radio emission on such high harmonics by any mechanisms. We suppose that this emission is generated at low cyclotron harmonics, by some energetic electron beam, in significantly inhomogeneous magnetic field. As examples we consider two microwave bursts with the fine structures, which were observed with the Beijing Astronomical Observatory spectrometer (China) and the radiospectrograph at the Ondrejov Observatory (Czech Republic).

\section[]{Observations}

The observations were performed by the newly developed Solar Radio Broadband Fast Dynamic Spectrometers (1-2, 2.6-3.8, and 5.2-7.6 GHz frequency range) at Huairou Solar Observing Station of the Beijing Astronomical Observatory (Fu et al., 1995), and by the 1-4.5 GHz radiospectrograph at the Ondrejov Observatory (Jiricka et al., 1993).

\subsection{The April 15, 1998 event}

 In April 15, 1998 during a weak solar flare (SN, 07:39-08:06 UT), which occurred in the active region NOAA 8203 at position N29W15, the radio bursts were observed. There are many complex radio fine structures in the time interval 07:59:30-08:01:40 UT, e.g. the sub-second type U burst and millisecond spikes, registered by the 1.0-2.0 GHz and 2.6-3.8 GHz spectrometers at BAO (Huang and Fu, 1999; Wang et al., 2000).
     The event discussed here was recorded by the 2.6-3.8 GHz spectrometer and it is shown in Figure 1 as the cascade plot of radio spectrograms in the right polarization. The left polarization was very weak. So it was a strong polarized event. In the time interval 08:01:27-08:01:31 UT the fine structure consisting of three narrow frequency bands ($\sim 20$ MHz) was observed.  The average distance between the bands is $\sim 50$ MHz. Especially distinct bands were observed at 08:01:29-08:01:30 UT. At this time all bands show drift to higher frequencies (df/dt $\sim300$ MHz $s^{-1}$), and then this drift changes to lower ones (df/dt $\sim -100$ MHz $s^{-1}$). 

\subsection[]{The July 12, 2000 event}

During the July 12, 2000 flare (X1.9/2B, 10:18-10:46 UT, N17E27, in the NOAA 9077 active region), at 10:35:45-10:36:45 UT the unique drifting zebra pattern was observed by the 1-4.5 GHz Ondrejov radiospectrograph  (Figure 2). It consists of two branches which frequencies are in harmonic ratio (1:2): a) the fundamental frequency branch, and b) the harmonic frequency one.  In both branches the fine structures, superimposed on the weak narrowband drifting continua, can be seen. The bandwidth of the continuum is measurable better in the harmonic branch ($\sim 400$ MHz). Both branches are drifting towards lower frequencies. The mean frequency drift of the harmonic branch is -15 MHz $s^{-1}$, in the 10:36:00-10:36:30 UT interval. The maximum drift ($\sim$ -30 MHz $s^{-1}$) in the harmonic branch was observed during the observation of the zebra pattern at 10:36:15-10:36:22 UT. The zebra pattern in the harmonic branch consists of 4 nearly parallel drifting narrowband ($\sim 50$ MHz) lines. Parameters of this zebra pattern are summarized in Table 1. It is important to mention that the ratio of the zebra line frequencies is decreasing with the frequency decrease, with one exception. Furthermore, in the harmonic branch at 10:36:00 and 10:36:30 UT weak zebra patterns can be recognized, but by a special inspection. On the other hand the fundamental branch shows also fine structures, in some sense similar and also different. For example, at time (10:36:15-10:36:22 UT) of the zebra pattern we can see in the 1.15-1.3 GHz frequency range a weak, but enhanced continuum, in which two weak zebra line can be recognized. On the other hand, at 10:35:53-10:36:14 UT (the time of the very weak zebra pattern in the harmonic branch), in the 0.8-1.45 GHz the drifting structure was observed. This structure consists of several lines, some of them are splitted in frequency. It looks that some part of this structure has a counterpart in the harmonic branch, but some part, especially on lower frequencies (0.8-1.1 GHz), not. The frequency drift of the narrowband line (starting at 10:35:53 UT, 1.43 GHz and ending at 10:36:14 UT, 1.15 GHz) is -13 MHz $s^{-1}$, and its bandwidth is$ \sim 20$ MHz.

\section{Estimation of the magnetic field strength}

If an electron beam moves in plasma with the magnetic field, it excites longitudinal waves at the Cherenkov resonance and anomalous Doppler effect resonance (Mikhailovsky, 1975). A spectrum of the waves determines a spectrum of solar radio emission, which originates due to a conversion of the longitudinal waves into electromagnetic ones (Zheleznyakov, 1977). But the bandwidth of the emission at the Cherenkov resonance is determined mainly by the length of the beam, which is relatively large and, as observations show, cannot be less than about 10 \% without special suppositions concerning an inhomogeneity of plasma or short length of the beam. One can see such emission on Figure 1 as the fast drifting structures which intensity is changing mainly in the course of time. The emission at the anomalous Doppler effect has also relatively large bandwidth, because the resonance condition $\omega - k_{\parallel} v_{\parallel} + \omega_{Be} = 0$ can be satisfied in the wide range of $k_{\parallel}$ values and the bandwidth of the longitudinal waves spectra is determined by the length of the beam also. Here $\omega$ is the longitudinal wave frequency, $k_{\parallel}$ is the component of the wave vector along the magnetic field, $v_{\parallel}$ is the component of the electron velocity along the magnetic field and $\omega_{Be}$ is the electron cyclotron frequency. 

The only one possibility to explain the narrow band emission, if do not special suppositions concerning the inhomogeneity of the plasma and parameters of the beam, is the resonance at the normal Doppler effect, namely, $\omega - k_{\parallel} v_{\parallel} - s\omega_{Be} = 0$ (s is the cyclotron harmonic number) on the anisotropic distribution function of energetic electrons, when $T_{\perp} > T_{\parallel}$ and $k_{\parallel} v_{\parallel} < \omega_{Be}$. Here $T_{\perp}$ and $T_{\parallel}$ are the temperatures of energetic electrons across and along the magnetic field, respectively. This resonance condition is fulfilled at the fixed level in the corona under an assumption that a transverse dimension of the electron beam is small enough such that the density of the background plasma does not change more than few percents on the transverse dimension of the beam. The anisotropic distribution function is formed by a natural way due to differentiation in space of the electrons with different velocities. In this case the higher cyclotron harmonics can be excited; higher than in the case of the isotropic beam distribution function (Mikhailovsky, 1975). The condition $k_{\parallel} v_{\parallel} < \omega_{Be}$ is fulfilled, because the growth rate of the longitudinal waves is (Mikhailovsky, 1975)

\begin{displaymath}
\gamma =\frac{\sqrt{\pi }\alpha \omega ^3}{v_{Te\Vert }^3k^2|k_{\Vert }|}%
{\sum\limits_{s=-\infty }^\infty I_s\exp (-z_{\perp })\exp \left[ -(\frac{\\
\omega -k_{\Vert }v_{\Vert }-s\omega _{Be}}{k_{\Vert }v_{Te\Vert }}\\
)^2\right] \left[\omega-k_{\Vert}v_{\Vert}-s\omega_{Be}(1-\frac{T_{\Vert}}{T_{\perp}})\right]} 
\end{displaymath}

i.e. inversely proportional to $k_{\parallel}$. At the same time the value of $k_{\perp}$ is limited from below, because $\gamma \rightarrow 0$ with $k_{\perp} \rightarrow 0$ under $s > 2$ (under s = 2 we have $\gamma \rightarrow Const$ with $k_{\perp} \rightarrow 0$). Here $\alpha$ is the ratio of the beam density $n_{b}$ to the density of background plasma n, $\omega_{pe}$ is the electron plasma frequency, $v_{Te\parallel}$ is the thermal electron velocity along the magnetic field, k is the wave number, $I_{s} (k_{\perp} v_{\perp} / \omega_{Be})$ is the modified Bessel function, $z_{\perp} = k_{\perp}^2 T_{\perp} / m\omega_{Be}^2$, m is the electron mass, $k_{\perp}$ is the transverse component of the wave number. 

So in this case we have $k_{\parallel} << k_{\perp}$ for the most effectively excited longitudinal waves. It means that the frequency of excited waves is $\omega \approx (\omega_{pe}^2 + \omega_{Be}^2)^{1/2}$ and $\omega \approx s\omega_{Be}$, under the condition $k_{\parallel} v_{\parallel} < \omega_{Be}$ and for the case with the cold background plasma. It is also the favorable condition for an emission generation on the second harmonic of the longitudinal waves due to a mergence of them (Zheleznyakov, 1977). The generation of the fundamental emission (at the first harmonic of the longitudinal wave frequency) is possible, but there is the problem of the emission escape. From here one can see that the cyclotron harmonics are excited beginning from the second one.

Hence we can estimate the medium parameters in the region of the emission under suggestion that the emission takes place on the second harmonic of the longitudinal waves, i.e. $\omega^t \approx 2\omega$. Consequently, $\omega_{pe} \approx (\omega^t/2)(1 - 1/s^2)^{1/2}$ and $\omega_{Be} \approx \omega^t/2s$, where $\omega^t$ is the frequency of the radio emission. The problem is to find the harmonic number s for the determination of $\omega_{Be}$. For the determination of $\omega_{pe}$ it is not necessary to know s, because $\omega_{pe} \approx \omega^t/2$ for $s \geq 3$ with the accuracy a few percents. As for s = 2, this relation gives the estimation of $\omega_{pe}$ with accuracy up to 15 \%.

Hence we can estimate the ratio of the emission frequencies for adjacent harmonics, namely, $\omega_{s} /\omega_{s+1} = (\omega_{pes}/\omega_{pes+1})[s^3(s+2)/(s+1)^3(s-1)]^{1/2}$. Here $\omega_{s}$ and $\omega_{s+1}$ are the frequencies of the emission corresponding to s and s+1 cyclotron harmonics, $\omega_{pes}$ and $\omega_{pes+1}$ are the electron plasma frequencies corresponding to levels of the emission of the s and s+1 harmonics. If the density changes slowly with the height so that $\omega_{pe}$ changes slower than $\omega_{Be}$, we can take $\omega_{pes} \approx \omega_{pes+1}$ and consequently we have

$\omega_{s}/\omega_{s+1} \approx [s^3 (s+2)/(s+1)^3 (s-1)]^{1/2}$. 

Hence we have $\omega_{2}/\omega_{3} \approx 1.09$, $\omega_{3}/\omega_{4} \approx 1.03$ and $\omega_{4}/\omega_{5} \approx 1.01$. These are minimal ratios, because really $\omega_{pe}$ changes also.

It should be noted that we can not exclude the possibility of higher harmonics generation under such ratios of frequencies, if the magnetic field changes slowly or the density changes quickly enough with height. But in our opinion it is unlikely, because of the growth rate of the longitudinal waves and consequently the intensity of the emission decreases quickly with the increase of the harmonic number.  
Consequently in this case, when wpe changes essentially slower than wBe, we can distinguish only not more than four first harmonics, if the frequency band of the emission is about 1%. The rest of harmonics merges. If the scale height of wpe is comparable with the one of wBe we can observe more than four harmonics. One can see that the highest frequency corresponds to the lowest (second) harmonic. If we suppose that wBe changes slower than wpe, we have harmonic ratios of frequencies, and the sequence of harmonics changes, i.e. higher harmonics have higher frequencies.

On Figure 1 one can see the spectrogram with three frequency bands. Their average ratios are not lower than 1.03. So we can conclude that there are three harmonics beginning from the third one on the Figure 1. The frequency of the third cyclotron harmonic is about 1.8 GHz and the corresponding magnetic field at the level of third harmonic emission is $\sim 200$ G.
On Figure 2 there are four frequency bands well distinguishable. It makes possible to determine the ratios of the frequencies of adjacent frequency bands at different moments of time accurately enough. The results are shown in Table 1.  The ratios of lowest adjacent frequency bands are more than 1.03 and less than 1.09. It means that we have the cyclotron harmonic emission beginning from the third one. The frequency of the third cyclotron harmonic is 1.3 GHz and the corresponding magnetic field strength is about 140 G. 

It is essential that in the event on July 12, 2000 the fundamental emission has observed. Its intensity is lower than the one at the second harmonic of the longitudinal wave frequency due to strong collision damping (Benz, 1993) and its structure is different apparently due to another mechanism of the longitudinal waves conversion into the radio emission.

\section{Estimation of density variations}

During a non-stationary process the ratio between the density and the magnetic field changes. If a coronal mass ejection takes place, the magnetic field gradient may decrease due to a drawing out of the magnetic field lines. Hence the ratio between adjacent cyclotron harmonics increases. First of all, it increases at higher harmonics, because the magnetic field is drawn out more significantly at larger heights in the corona. Thus, the frequency distances among the harmonic frequency bands can become equal. Such event can be seen on Figure 1 at 08:01:28.5 UT. There is a small coronal ejection with duration about 0.1 s. The emission frequency decrease is ~ 1.5%. Since in our case the kinetic pressure is much less than the magnetic pressure, then under any disturbances the relative variation of the magnetic field is much less than the corresponding variation of the density under supposition that the temperature does not change. It means that the frequency decrease of ~ 1.5 % corresponds to the density decrease in the emission region of ~ 3%. At 08:01:29 UT (Figure 1) there is the increase of the frequency emission of ~ 4% during the time ~ 0.5 s. This corresponds to the density growth of ~ 8%. 

\section{Conclusions}

Observations of the solar radio emission in the microwave range are made at Beijing Astronomical Observatory (China) and Ondrejov Observatory (Czech Republic) with the spectrographs of high temporal and spectral resolutions. These observations show different fine structures superimposed on the microwave bursts. In particular, there are the observed structures consisted of almost parallel narrow frequency bands (zebras) with the ratios of the adjacent bands frequencies not more than 1.05.

For an interpretation of these events the model of the emission generation by the anisotropic electron beam at low cyclotron harmonics is suggested. Such an electron beam excites longitudinal waves at the normal Doppler effect resonance. Then these waves are converted into the radio emission at the second harmonic of the longitudinal wave frequency and sometimes also into the fundamental emission. This model gives the minimal values of the cyclotron harmonic ratios: $(\omega_{2}/\omega_{3}) = 1.09$, $(\omega_{3}/\omega_{4}) = 1.03$, $(\omega_{4}/\omega_{5}) = 1.01$. Using these values one can determine the numbers of the cyclotron harmonics corresponding to each frequency bands and then determine the magnetic field strength. For events observed at Beijing Astronomical Observatory and Ondrejov Observatory we determined that emissions took place at the cyclotron harmonics beginning from third one. Hence we determined the magnetic field strength $\sim 200$ G for the emission, which observed at the Beijing Astronomical Observatory and $\sim 140$ G for that observed at the Ondrejov Observatory.

\begin{table}[p]

\caption{The parameters of zebra pattern}

\begin{tabular}{cccc}

Time [UT]  & Line No  &  Frequency [MHz]  & Frequency ratio\\

10:36:18    &   3           &    2375 \\
                 &                 &                               &     1.034 \\
                  &   2           &    2455 \\
                  &                &                               &     1.046 \\
                  &   1           &    2568 \\
        
10:36:19    &   4            &    2284 \\
                  &                 &                               &     1.025 \\
                  &   3            &    2341 \\
                  &                 &                               &     1.044 \\
                  &   2            &    2443 \\
                  &                 &                               &     1.040 \\
                  &   1            &    2541 \\

10:36:20    &   4             &    2250 \\
                 &                   &                              &     1.025 \\
                 &   3              &    2307 \\
                 &                   &                              &     1.039 \\
                 &   2              &    2398 \\
                 &                   &                              &     1.047 \\
                 &   1              &    2511 \\ 

10:36:21    &   2              &    2364 \\
                 &                   &                              &     1.048 \\
                 &   1              &    2477 \\

\end{tabular}
\end{table}
                                                             
\begin{figure}[t]
\plotone_reduction{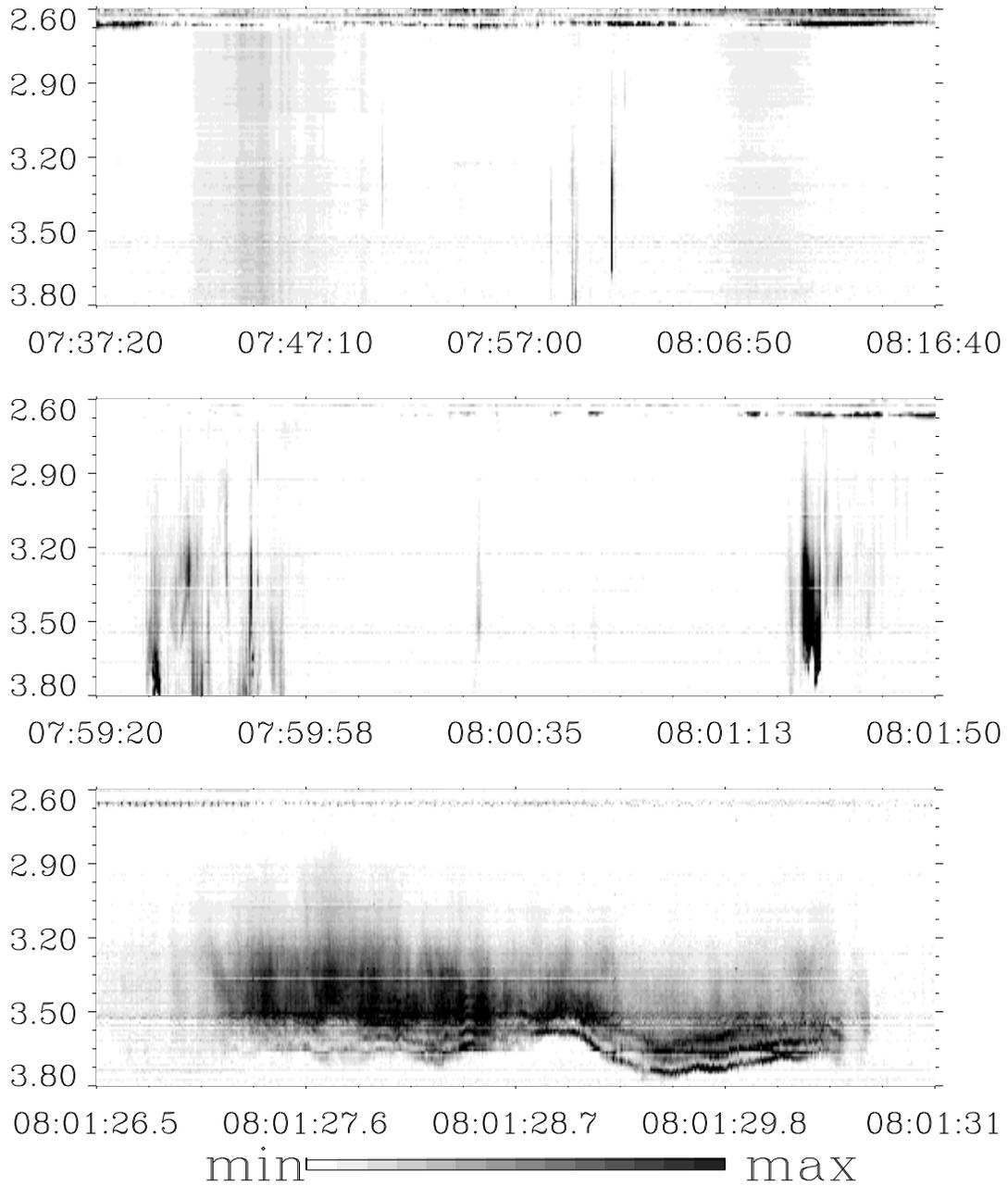}{1.0}
\caption{The cascade plot of the radio spectrograms in the right polarization, showing the fine structures of the solar bursts on 15 April 1998 in the 2.6-3.8 GHz range observed with 10 MHz spectral resolution at BAO. From top to bottom the temporal resolutions are 0.2 s, 0.2 s, and 8 ms, respectively. The intensity in arbitrary units is shown in gray scale.}
\end{figure}

\begin{figure}
\plotone_reduction{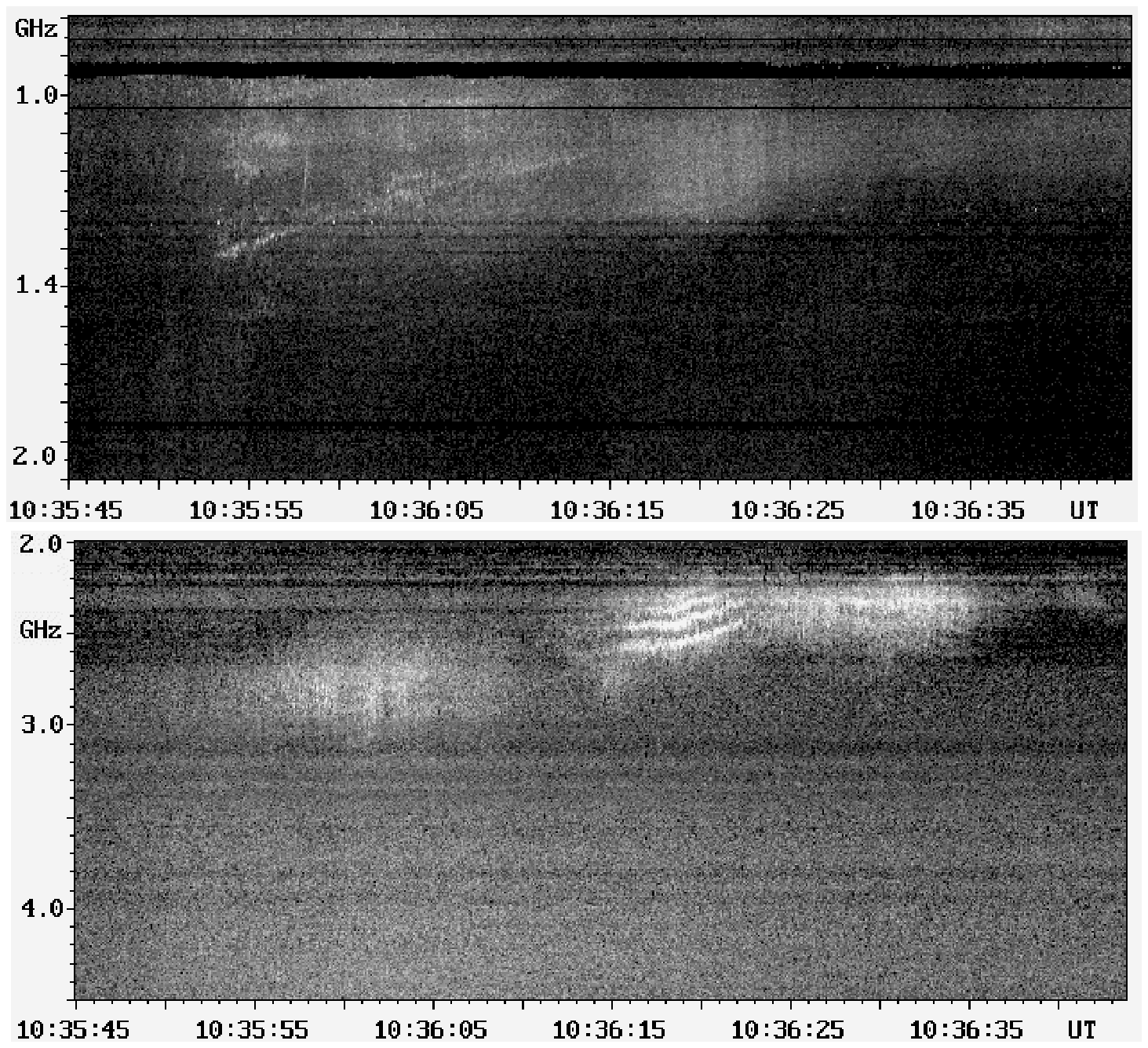}{1.0}
\caption{The radio spectrum of the drifting zebra pattern observed during the July 12, 2000 flare by the Ondrejov radio spectrograph.}
\end{figure}

\bigskip

\begin{bf}
Acknowledgment
\end{bf}
The work is supported by the Chinese Academy of Sciences and the NSFC Grants: 19773016, 19833050, 49990451 and Russian Basic Research Foundation (Grant 98-02-17727). We acknowledge the Huairou staff members for operating the Radio Spectrographs properly. 

\bigskip
{\small
\begin{center}
{\bf REFERENCES}
\end{center}

\begin{description}
\itemsep -0.6mm
\item Dulk, G., McLean, D.J., 1977,  Solar Physics, 57, 279. 
\item Gelfreikh, G.B., 1998,  Astron. Soc. Pac. Conf. Ser., 155, 110. 
\item Huang, G., and Fu, Q., 1999, in Solar Physics with Radio Observations, Proc. Nobeyama Symp.,1998, NRO Report No. 479, p.287.
\item Isliker, H., and Benz,A.O., 1994, Astron. Astrophys. Suppl. Ser., 104, 145.
\item Ji, H., Fu, Q., Liu, Y., Cheng, C., Chen, Z., Lao, D., Ni, C., Pei, L., Xu, Z., Chen, S., Yao, Q., Qin, Z. and Yang, G., 2000, ACTA Astrophysica Sinica, 20, 209.
\item Jiricka K., Karlicky M., Kepka O., Tlamicha A., 1993, Solar Phys.,147, 203.
\item Karlicky M., Jiricka K., 1995, Solar Phys., 160,121.
\item Karlicky M., Sobotka M., Jiricka K., 1996, Solar Phys., 168, 375.
\item Ledenev, V.G. and Messerotti, M., 1999, Solar Phys., 185, 193.
\item Mann G., Karlicky M., Motschmann U., 1987, Solar Phys., 110, 381.
\item Mikhaylovsky, A.B., 1975, Theory of Plasma Instabilities, Nauka, Moscow, Vol.1.
\item Wang, S., Yan, Y. and Fu, Q., 2000, in Heating and Energetics of the Solar Corona and Solar Wind, COSPAR 33 SA, Warsaw, Poland.
\item Zheleznyakov, V.V., 1977, Electromagnetic Waves in Space Plasma, Nauka, Moscow. 
\end{description}

\end{document}